\documentclass{article}
\usepackage{spconf,amsmath,epsfig}

\usepackage{times}
\usepackage{epsfig}
\usepackage{graphicx}
\usepackage{amsmath}
\usepackage{amssymb}

\usepackage{dsfont}
\usepackage{booktabs}
\usepackage{pifont}
\usepackage{arydshln}
\usepackage{multirow}
\usepackage[ruled]{algorithm2e}
\SetKwComment{Comment}{$\triangleright$\ }{}
\usepackage{url}            % simple URL typesetting

\usepackage{xcolor}

\usepackage{soul}

\newcommand{\ie}{\textit{i}.\textit{e}.}

\let\OLDthebibliography\thebibliography
\renewcommand\thebibliography[1]{
  \OLDthebibliography{#1}
  \setlength{\parskip}{0pt}
  \setlength{\itemsep}{0pt plus 0.3ex}
}

\begin{document}\sloppy

% Example definitions.
% --------------------
\def\x{{\mathbf x}}
\def\L{{\cal L}}

% Title.
% ------
\title{Towards Robust Data Hiding against (JPEG) Compression: \\
A Pseudo-differentiable Deep Learning Approach}
%
% Single address.
% ---------------
\name{Chaoning Zhang\thanks{$^*$Equal contribution}$^*$, Adil Karjauv$^*$, Philipp Benz$^*$, In So Kweon}
%Address and e-mail should NOT be added in the submission paper. They should be present only in the camera ready paper. 
\address{\small \texttt{chaoningzhang1990@gmail.com}, \texttt{mikolez@gmail.com}, \texttt{pbenz@kaist.ac.kr}, \texttt{iskweon77@kaist.ac.kr}}

\maketitle

\begin{abstract}
Data hiding is one widely used approach for protecting authentication and ownership. Most multimedia content like images and videos are transmitted or saved in the compressed form. This kind of lossy compression, such as JPEG, can destroy the hidden data, which raises the need of robust data hiding. It is still an open challenge to  achieve the goal of data hiding that can be against these compressions. Recently, deep learning has shown large success in data hiding, while non-differentiability of JPEG makes it challenging to train a deep pipeline for improving robustness against lossy compression. The existing SOTA approaches replace the non-differentiable parts with differentiable modules that perform similar operations. Multiple limitations exist: (a) large engineering effort; (b) requiring a white-box knowledge of compression attacks; (c) only works for simple compression like JPEG. In this work, we propose a simple yet effective approach to address all the above limitations at once. Beyond JPEG, our approach has been shown to improve robustness against various image and video lossy compression algorithms. Code: \url{https://github.com/mikolez/Robust_JPEG}.
\end{abstract}


\section{Introduction}
The Internet has revolutionized the world development in the sense that it has become the de facto most convenient and cost-effective remote communication medium~\cite{bao2018robust}. Early internet users mainly communicated in the form of plain text, and in recent years, images or even videos have gradually become the dominant multimedia content for communication. However, those images are vulnerable to security attacks including unauthorized modification and sharing. 
Data hiding has been established as a promising and competitive approach for authentication and ownership protection~\cite{alenizi2017robust}.

Social media has become indispensable part for most of us who are active on the FaceBook, Twitter, Youtube for sharing multimedia content like photos or videos. For reducing the bandwidth or traffic to faciliate the transmission, most photos and videos on the social media (or internet in general) are in the compressed form, such as JPEG or MPEG. This kind of lossy operation can easily destroy the hidden information for most existing data hiding approaches~\cite{bao2018robust,alenizi2017robust}. Thus, robust data hiding in multimedia that can resist non-malicious attack, lossy compression, has become a hot research direction. 

Recently, deep learning has also shown large success in hiding data in images~\cite{hayes2017generating,weng2018convolutional,wu2018image,zhu2018hidden,ahmadi2020redmark}. For improving the robustness against JPEG compression, the SOTA approaches need to include the JPEG compression in the training process~\cite{zhu2018hidden,ahmadi2020redmark}. The challenge of this task is that JPEG compression is non-differentiable. Prior works have either roughly simulated the JPEG compression~\cite{zhu2018hidden} or carefully designed the differentiable function to mimic every step of the JPEG compression~\cite{ahmadi2020redmark}. Compared with the rough simulation approach~\cite{zhu2018hidden}, the carefully designed mimicing approach has achieved significant performance~\cite{ahmadi2020redmark}. However, it still has several limitations: (a) mimicking the JPEG compression requires large amount of engineering effort, (b) it is compression specific, i.e. it only works for JPEG compression, while not effective against other compression, (c) it requires a full white-box knowledge of the attack method. In practice, the attacker might have some publicly available tool.
In this work, we demonstrate that a simple approach can be adopted for alleviating all the above limitations at once.

\section{Related work}
\subsection{Lossy Data compression}
In information technology, data compression is widely used to reduce the data size~\cite{sayood2017introduction}. The data compression techniques can be mainly divided into lossless and lossy ones~\cite{sayood2017introduction}. Lossless techniques have no distortion effect on the encoded image, thus it is not relevant to robust data hiding. We mainly summarize the lossy compression for images and videos. Overall, lossy compression is a class of data encoding methods that discards partial data. Without noticeable artifacts, it can reduce the data size significantly, which facilitates storing and transmitting digital images. 
JPEG was first introduced in 1992 and has become the de facto most widely used compression technique for images~\cite{pennebaker1992jpeg}. JPEG compression involves two major steps: DCT transform and quantization. For color images, color transform is also often adopted as a standard pre-processing step~\cite{pennebaker1992jpeg}. Even though humans cannot easily identify the difference between original images and compressed images, this lossy compression can have significant influence on the deep neural networks.  JPEG2000~\cite{skodras2001jpeg} is another famous lossy compression that has higher visual quality with less mosaic artifacts. In contrast to JPEG, JPEG2000 is based on Discrete Wavelet Transform~\cite{lu2009robust}. Concurrent to JPEG2000, another variant of DWT based lossy compression, \ie\ progressive file format Progressive Graphics File (PGF)~\cite{stamm2002new} has also come out. In 2010, Google developed a new image format called WebP, which is also based on DCT as JPEG but outperforms JPEG in terms of compression ratio~\cite{ginesu2012objective}. For video lossy compression, MPEG~\cite{le1991mpeg} is the most widely adopted approach.

\subsection{Traditional robust data hiding}
Since the advent of image watermarking, numerous works have investigated robust data hiding to improve its imperceptibility, robustness as well as hiding capacity. Early works have mainly explored manipulating the cover image pixels directly, \ie\ embedding the hidden data in the spatial domain, for which hiding data in least significant bits (LSBs) can be seen as a classical example. For improving robustness, the trend has shifted to hiding in a transform domain, such as Discrete Cosine Transform (DCT)~\cite{singh2018new}, Discrete Wavelet Transform~\cite{lu2009robust}, or a mixture of them~\cite{singh2012robust,harish2013hybrid}. For achieving a good balance between imperceptibility and robustness, adaptive data hiding has emerged to embed the hidden data based on the cover image local features~\cite{bhinder2018image,chen2018quaternion}. Traditional data hiding methods often require the practitioners to have a good understanding of a wide range of relevant expertise. Deep learning based approaches automated the whole process, which significantly facilitates its wide use. Moreover, recent works have demonstrated that deep learning based approaches have achieved competitive performance against traditional approaches in terms of capacity, robustness, and security~\cite{zhu2018hidden,ahmadi2020redmark}. 

\subsection{Deep learning based robust data hiding}
Beyond the success in a wide range of applications, deep learning has shown success in data hiding~\cite{isac2011study}. The trend has shifted from adopting DNNs for a certain stage of the whole data hiding pipeline~\cite{jin2007applications,kandi2017exploring,mun2017robust} to utilizing DNNs for the whole pipeline in an end-to-end manner~\cite{hayes2017generating}. ~\cite{baluja2017hiding} has shown the possibility of hiding a full image in a image, which has been extended to hiding video in video in~\cite{weng2018convolutional,wu2018image}. Hiding binary data with adversarial learning has also been proposed in~\cite{hayes2017generating}. The above deep learning approaches do not take robustness into account and the hidden data can easily get destroyed by common lossy compression, such as JPEG. For improving its robustness to JPEG, ~\cite{zhu2018hidden} have proposed to include ``noise layer" that simulates the JPEG to augment the encoded images in the training. To overcome the challenge that ``true" JPEG is non-differentiable, the authors propose to ``differentiate" the JPEG compression with two approximation methods, i.e. JPEG-Mask and JPEG-Drop, to remove the high frequency content in the images. We refer the readers to ~\cite{zhu2018hidden} for more details. Even though JPEG-Mask and JPEG-Drop also remove the high frequency content as the ``true" JPEG, their behavior is still quite far from the ``true" JPEG, resulting in over-fitting and poor performance when tested under the ``true" JPEG. To minimize the gap approximation and ``true" JPEG, more recently, ~\cite{ahmadi2020redmark} proposed a new approach to carefully simulate the important steps in the ``true" JPEG. Similar approach for simulating JPEG has also been proposed in~\cite{shin2017jpeg} for generating JPEG-resistant adversarial examples.

\begin{figure*}[t]
    \centering
    \scalebox{0.9}{
    \includegraphics[width=\linewidth]{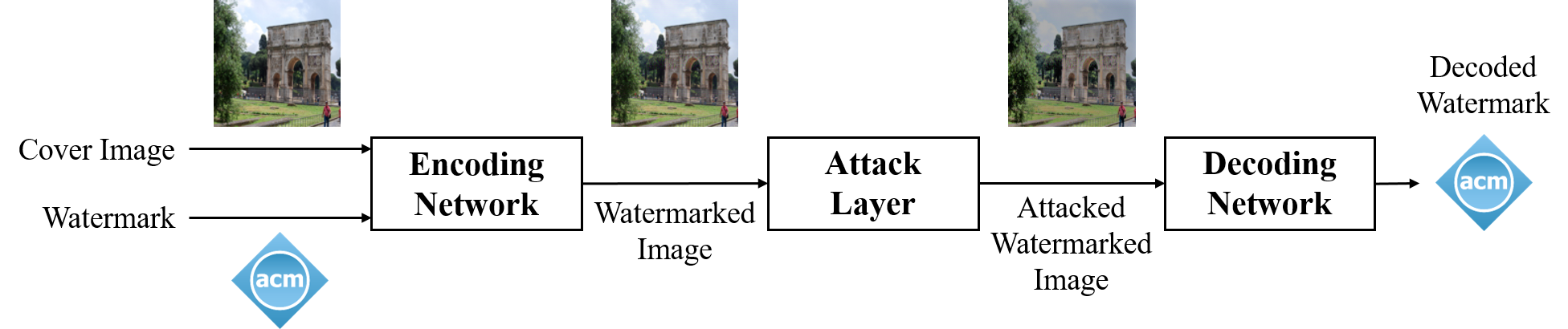}
    }
    \caption{Main framework.}
    \label{fig:main_framework}
\vspace{-2mm}
\end{figure*}

\section{Pseudo-Differentiable JPEG and Beyond}
Robust data hiding requires imperceptibility and robustness. The imperceptibility is achieved minimizing its distortion on the cover image. In this work, we aim to improve robustness against lossy compression, such as JPEG, while maintaining its imperceptibility. Specifically, we focus on deep learning based approach through addressing the challenge that lossy compression is non-differentiable. Note that lossy compression by default is always non-differentiable due to many factors, out of which quantization is one of the widely known factors. 

\subsection{Why do we focus on lossy compression?} 
A intentional attacker can adopt various techniques, such as nosing, filtering or compression, to remove the hidden data. In practice, however, unauthorized sharing or modication is often done without any intentional attacks. Instead, the social media platform often performs non-intentional attack through lossy compression, while it is very unusual that a media platform would intentionally add noise to the images. Thus, addressing the robustness against lossy compression has high practical relevance. Moreover, the existing approaches already have a good solution to other methods. 

\subsection{Deep Learning based robust data hiding SOTA framework}
Before introducing the proposed method for improving robustness against lossy compression, we first summarize the deep learning based robust data hiding framework adopted by SOTA approaches~\cite{zhu2018hidden,ahmadi2020redmark} in Fig~\ref{fig:main_framework}. Straightforwardly, it has two major components: encoding network for hiding the data in the cover image and decoding network for extracting the hidden data from the encoded image. Additionally it has an auxiliary component, \ie\ attack layer (also called as ``noise layer" in~\cite{zhu2018hidden}), necessary for improving its robustness against attacks, such as lossy compression. 

Our key contribution lies in \textit{providing a unified perspective on this attack layer and proposing a frustratingly simple technique for improving robustness against lossy compression}. 

\subsection{Pseudo differentiable lossy compression}
\begin{figure}[!htbp]
    \centering
    \scalebox{0.9}{
    \includegraphics[width=\linewidth]{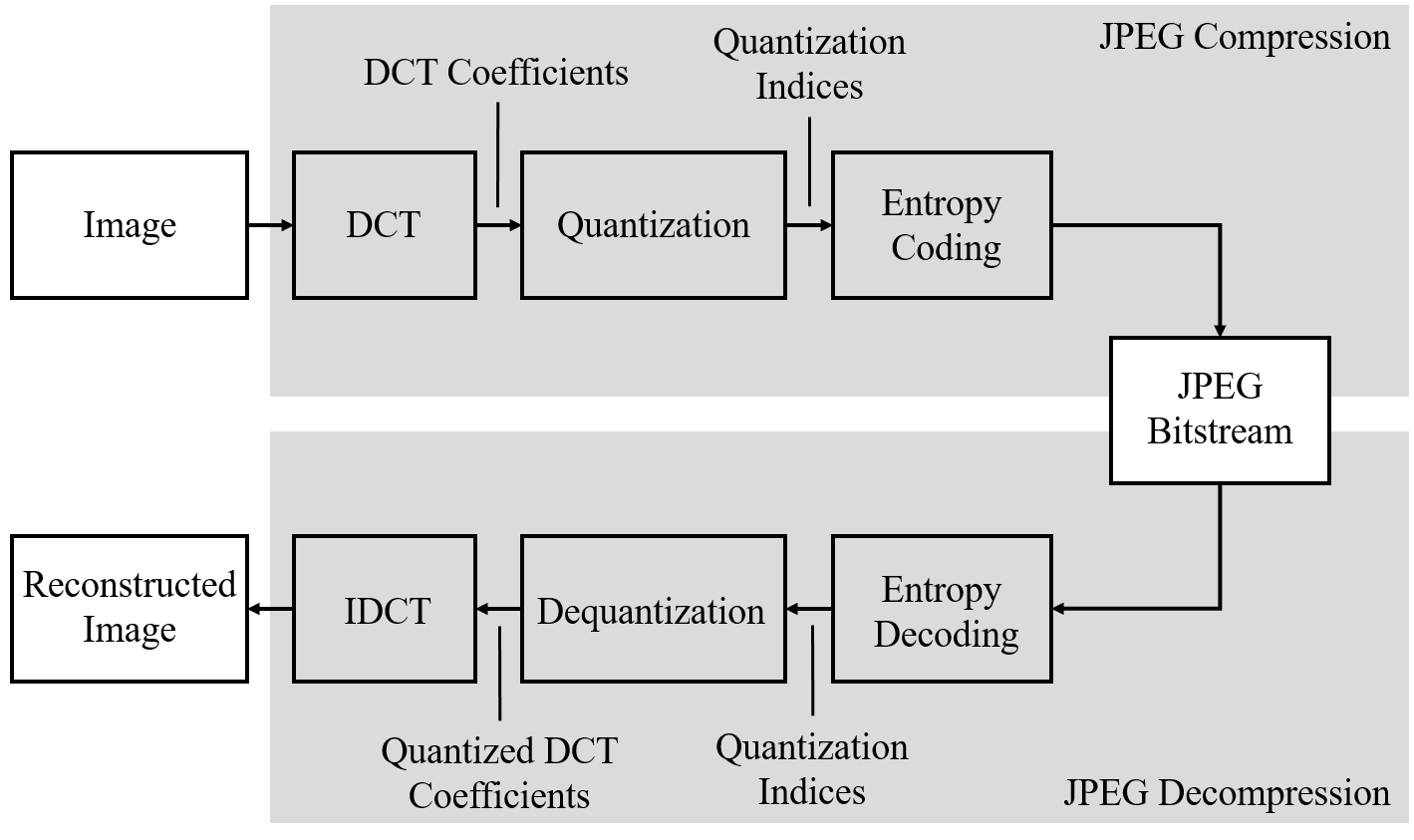}
    }
    \caption{JPEG compression and decompression.}
    \label{fig:jpeg_compression_decompression}
\vspace{-2mm}
\end{figure}

For general noise, such as Gaussian noise, the noise can be directly added to the encoded image. The role of such noise is to augment the encoded images in the training process to make it robust against such noise. Note that the operation of adding noise is differentiable, thus it is fully compatible with the need of being differentiable for the attack layer. Other attack operations like filtering are also differentiable. The process of JPEG compression and decompression is shown in Fig.~\ref{fig:jpeg_compression_decompression}, where quantization involves rounding operation which is non-differentiable. It renders the JPEG unfit for gradient-based optimization~\cite{zhu2018hidden} and it constitutes a fundamental barrier for directly applying it in the attack layer of Fig.~\ref{fig:main_framework}. 

\textbf{Unified perspective on attack layer.} Regardless of the attack type, the resulting effect is a distortion on the encoded image. Specifically, Gaussian noise is independent of the encoded image while the JPEG distortion can be seen as a pseudo-noise that is dependent on the encoded image. 
\begin{figure}[!htbp]
    \centering
    \scalebox{1.0}{
    \includegraphics[width=\linewidth]{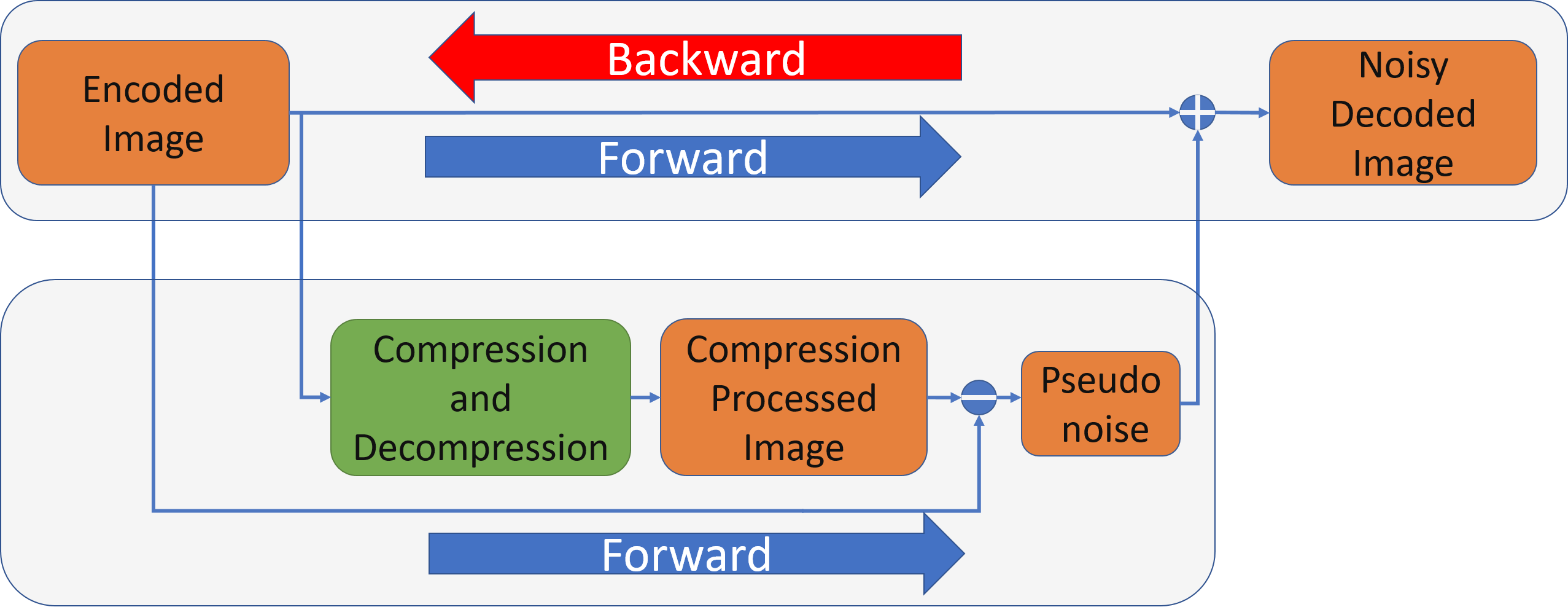}
    }
    \caption{Proposed pseudo-differentiable compression method.}
    \label{fig:pseudo_jpeg}
\vspace{-2mm}
\end{figure}

\begin{figure}[!htbp]
    \centering
    \scalebox{1.0}{
    \includegraphics[width=\linewidth]{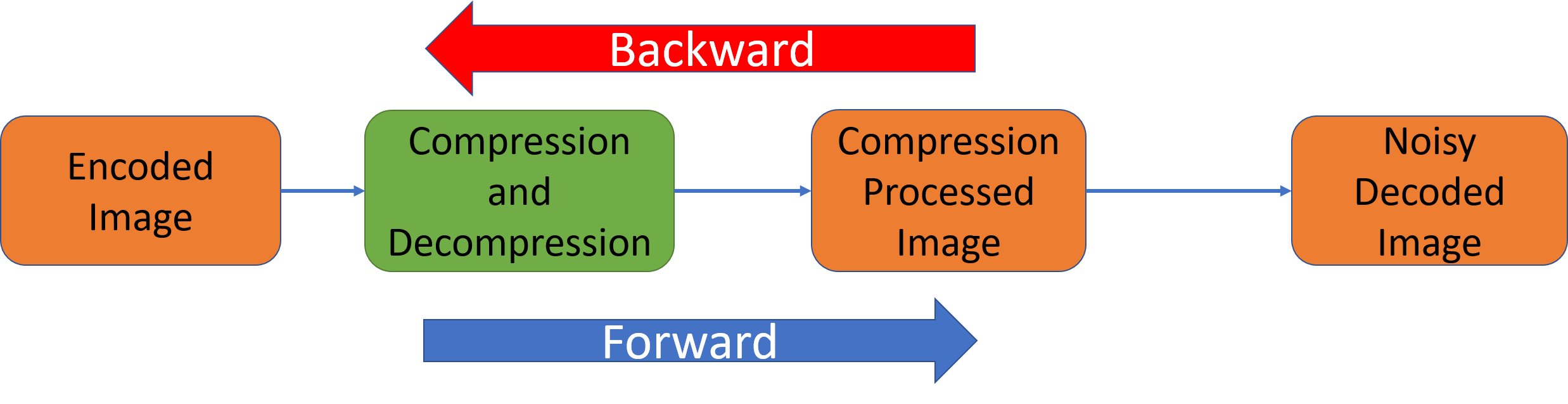}
    }
    \caption{Proposed pseudo-differentiable compression method.}
    \label{fig:true_jpeg}
\vspace{-2mm}
\end{figure}
\textbf{Pseudo-differentiable JPEG method.} Given the above unified perspective, we notice that the effect of the JPEG compression and decompression is equivalent to adding pseudo noise to the encoded image. Since adding Gaussian noise is a differentiable operation, the operation of adding JPEG pseudo-noise is also differentiable. Note that similar to Gaussian noise we can directly add JPEG pseudo-noise to the encoded image after the JPEG pseudo-noise is known. With this understanding, the problem remains how to get the JPEG pseudo-noise. Recall that JPEG pseudo-noise is the gap between the compression processed image and original encoded image. As the output of the encoding network, the encoded image is readily available and the compression processed image can also be simply retrieved by performing the forward operation of JPEG compression and decompression to the encoded image. The whole process of our proposed Pseudo-differentiable method is shown in Fig.~\ref{fig:pseudo_jpeg}. Note that during training the forward operation is performed on both encoded image pathway and pseudo-noise pathway, while the backward operation is only performed on the encoded image pathway. Since the backward operation does not go through the pseudo-noise path, the fact that JPEG compression is non-differentiable does not make the training fail. Astute readers can find that compression processed image and noisy deconded image in Fig.~\ref{fig:pseudo_jpeg} have exactly identical values. At first sight it appears that substracting the pseudo-noise from the compression image and then re-adding it to the encoded image are two meaningless operations that offset each other. However, from the gradient-optimization perspective, they are not dummy operations but instead they are the key to make non-differentiable JPEG operation pseudo-differentiable. Romoving these two seemingly meaniless operation results in a process shown in Fig.~\ref{fig:true_jpeg} which is impractical because the backward operation goes through the non-differentiable JPEG compression.

\textbf{Beyond JPEG compression.} The above proposed approach can be used for any lossy compression or any non-differentiable operation. With our proposed approach, we treat the non-differentiable operation as a blackbox, and we only need to perform the forward operation. This is conceivably significant advantage because some non-differentiable operation might be too complex for the practioner to understand how it works and mimicing its every step as in~\cite{ahmadi2020redmark}.

\section{Results}
Following~\cite{zhu2018hidden}, we hide 30 bits of random binary data in 128x128 color images. The comparison with existing various JPEG approximation methods are shown in the Table~\ref{tab:jpeg_comparison} with the decoding bit error rate (BER) as the metric. Our method outperforms all existing techniques for different JPEG quality factors. 

\begin{table}[!htbp]
\small
    \caption{Comparison of our method with other JPEG approximation techniques. The results are reported with the metric of BER ($\%$).}
      \centering
      \setlength\tabcolsep{5pt}
      \scalebox{0.7}{
        \begin{tabular}{cccccc}
        \toprule
        JPEG Quality & JPEG-Drop & JPEG-Mask & JPEG-Drop + JPEG-Mask & JPEG~\cite{shin2017jpeg} & Ours \\
        \midrule
        JPEG-10 & $49.33$ & $46.23$ & $46.38$ & $35.68$ & $\mathbf{31.22}$ \\
        JPEG-25 & $48.50$ & $39.48$ & $41.38$ & $22.00$ & $\mathbf{16.75}$ \\
        JPEG-50 & $47.62$ & $32.77$ & $38.48$ & $15.85$ & $\mathbf{12.63}$ \\
        JPEG-75 & $46.75$ & $27.22$ & $37.10$ & $14.03$ & $\mathbf{12.41}$ \\
        JPEG-90 & $46.08$ & $18.57$ & $34.93$ & $13.81$ & $\mathbf{12.00}$ \\
        \bottomrule
    \end{tabular}
    }
    \label{tab:jpeg_comparison}
\end{table}

To our knowledge, there is no existing deep learning based approach that is robust against other types of lossy compression, such as JPEG-2000 and WebP. Our approach can be readily extended to them and the results are shown in the Table~\ref{tab:jpeg-2000} and Table~\ref{tab:webp}. We train the network with quality factor of 500. For evaluation, when the quality factor is higher than 500, the decoding BER is very low, showing our approach can also improve the robustness against JPEG2000. It is expected that the BER increases when the evaluation quality factor decreases. Similar trend can be observed for WebP where the training quality factor is set to 50. 

\begin{table}[!htbp]
\small
    \caption{Results for different quality levels of JPEG 2000.}
      \centering
      \setlength\tabcolsep{5pt}
      \scalebox{1}{
        \begin{tabular}{ccccccc}
        \toprule
        JPEG 2000 & 100 & 250 & 500 & 750 & 900 \\
        \midrule
        Ours & $35.29$ & $4.16$ & $00.31$ & $ 00.04 $ & $00.03$ \\
        \bottomrule
    \end{tabular}
    }
    \label{tab:jpeg-2000}
\end{table}

\begin{table}[!htbp]
\small
    \caption{Results for different quality levels of WebP.}
      \centering
      \setlength\tabcolsep{5pt}
      \scalebox{1}{
        \begin{tabular}{ccccccc}
        \toprule
        WebP & 10 & 25 & 50 & 75 & 90 \\
        \midrule
        Ours & $20.69$ & $15.37$ & $12.86$ & $ 12.17 $ & $12.14$ \\
        \bottomrule
    \end{tabular}
    }
    \label{tab:webp}
\end{table}

We further extend our approach to two common video compression methods: MPEG and XVID. The results are available in Table~\ref{tab:video_codecs_results}. We observe that the BER with our approach is quite low for both video compression methods. Note that we still hide the same amount of binary data in each frame of the video. Our algorithm achieves low BER even in challenging scenario of video compression. Note that without approach, the BER is close to 50\%, \ie\ random guess.

\begin{table}[!htbp]
\small
    \caption{Results for video compression methods.}
      \centering
      \setlength\tabcolsep{8pt}
      \scalebox{1}{
        \begin{tabular}{cccccc}
        \toprule
        MPEG4 & XVID \\
        \midrule
        $8.45$ & $13.59$ \\
        \bottomrule
    \end{tabular}
    }
    \label{tab:video_codecs_results}
\end{table}

\begin{figure}[!htbp]
    \centering
    \scalebox{0.9}{
    \includegraphics[width=\linewidth]{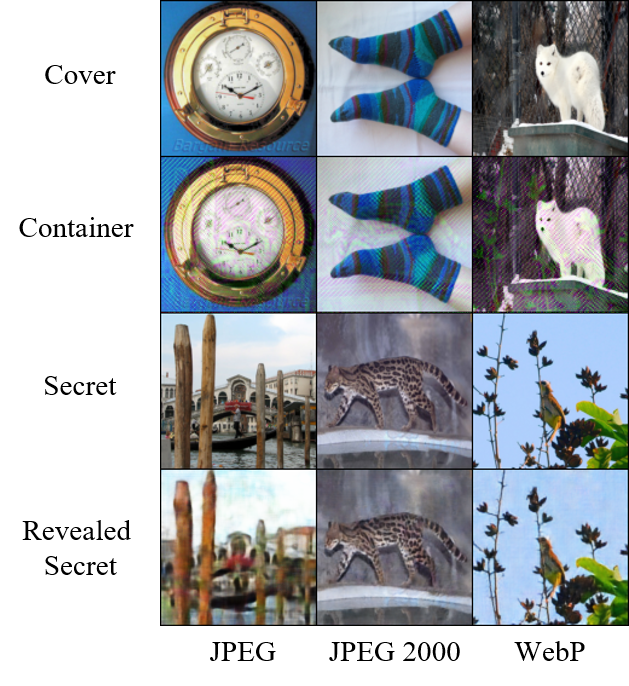}
    }
    \caption{Hiding an image in another under different lossy compressions.}
    \label{fig:img_diff_distortions}
\vspace{-2mm}
\end{figure}

Finally, we train a network to be simultaneously robust against JPEG, JPEG2000 and WebP by hiding an image in another. The average pixel discrepancy (APD) between cover and container images is 9.13/255, while that between secret and revealed secret images is 8.54/255. The qualitative results are shown in the Figure~\ref{fig:img_diff_distortions}.

\section{Conclusion}
In this paper, we provide a unified perspective on various attacks and propose a simple yet effective approach for addressing the non-differentiability for lossy compression. Our approach outperforms the existing approximation approach by a visible margin. Moreover, our work is the first to show that a deep learning approach can be robust against JPEG2020 and WebP as well as two video compression algorithms. We also show hiding images in images while staying simultaneously robust against several lossy compression. 

% References should be produced using the bibtex program from suitable
% BiBTeX files (here: strings, refs, manuals). The IEEEbib.bst bibliography
% style file from IEEE produces unsorted bibliography list.
% -------------------------------------------------------------------------
\bibliographystyle{IEEEbib}
\bibliography{bib_mixed}

\begin{thebibliography}{10}

\bibitem{bao2018robust}
Z~Bao, X~Luo, Y~Zhang, C~Yang, and F~Liu,
\newblock ``A robust image steganography on resisting jpeg compression with no
  side information,''
\newblock {\em IETE Technical Review}, 2018.

\bibitem{alenizi2017robust}
Farhan~A Alenizi,
\newblock {\em Robust Data Hiding in Multimedia for Authentication and
  Ownership Protection},
\newblock Ph.D. thesis, 2017.

\bibitem{hayes2017generating}
Jamie Hayes and George Danezis,
\newblock ``Generating steganographic images via adversarial training,''
\newblock in {\em Advances in Neural Information Processing Systems (NeurIPS)},
  2017.

\bibitem{weng2018convolutional}
Xinyu Weng, Yongzhi Li, Lu~Chi, and Yadong Mu,
\newblock ``Convolutional video steganography with temporal residual
  modeling,''
\newblock {\em arXiv preprint arXiv:1806.02941}, 2018.

\bibitem{wu2018image}
Pin Wu, Yang Yang, and Xiaoqiang Li,
\newblock ``Image-into-image steganography using deep convolutional network,''
\newblock in {\em Pacific Rim Conference on Multimedia}, 2018.

\bibitem{zhu2018hidden}
Jiren Zhu, Russell Kaplan, Justin Johnson, and Li~Fei-Fei,
\newblock ``Hidden: Hiding data with deep networks,''
\newblock in {\em Proceedings of the European Conference on Computer Vision
  (ECCV)}, 2018.

\bibitem{ahmadi2020redmark}
Mahdi Ahmadi, Alireza Norouzi, Nader Karimi, Shadrokh Samavi, and Ali Emami,
\newblock ``Redmark: Framework for residual diffusion watermarking based on
  deep networks,''
\newblock {\em Expert Systems with Applications}, 2020.

\bibitem{sayood2017introduction}
Khalid Sayood,
\newblock {\em Introduction to data compression},
\newblock 2017.

\bibitem{pennebaker1992jpeg}
William~B Pennebaker and Joan~L Mitchell,
\newblock {\em JPEG: Still image data compression standard},
\newblock 1992.

\bibitem{skodras2001jpeg}
Athanassios Skodras, Charilaos Christopoulos, and Touradj Ebrahimi,
\newblock ``The jpeg 2000 still image compression standard,''
\newblock {\em IEEE Signal processing magazine}, 2001.

\bibitem{lu2009robust}
Wei Lu, Wei Sun, and Hongtao Lu,
\newblock ``Robust watermarking based on dwt and nonnegative matrix
  factorization,''
\newblock {\em Computers \& Electrical Engineering}, 2009.

\bibitem{stamm2002new}
Christoph Stamm,
\newblock ``A new progressive file format for lossy and lossless image
  compression,''
\newblock in {\em Proceedings of the International Conferences in Central
  Europe on Computer Graphics, Visualization and Computer Vision, Plzen, Czech
  Republic}, 2002.

\bibitem{ginesu2012objective}
Giaime Ginesu, Maurizio Pintus, and Daniele~D Giusto,
\newblock ``Objective assessment of the webp image coding algorithm,''
\newblock {\em Signal Processing: Image Communication}, 2012.

\bibitem{le1991mpeg}
Didier Le~Gall,
\newblock ``Mpeg: A video compression standard for multimedia applications,''
\newblock {\em Communications of the ACM}, vol. 34, no. 4, pp. 46--58, 1991.

\bibitem{singh2018new}
Satendra~Pal Singh and Gaurav Bhatnagar,
\newblock ``A new robust watermarking system in integer dct domain,''
\newblock {\em Journal of Visual Communication and Image Representation}, 2018.

\bibitem{singh2012robust}
Surya~Pratap Singh, Paresh Rawat, and Sudhir Agrawal,
\newblock ``A robust watermarking approach using dct-dwt,''
\newblock {\em International journal of emerging technology and advanced
  engineering}, 2012.

\bibitem{harish2013hybrid}
NJ~Harish, BBS Kumar, and Ashok Kusagur,
\newblock ``Hybrid robust watermarking techniques based on dwt, dct, and svd,''
\newblock {\em International Journal of Advanced Electrical and electronics
  engineering}, 2013.

\bibitem{bhinder2018image}
Preeti Bhinder, Kulbir Singh, and Neeru Jindal,
\newblock ``Image-adaptive watermarking using maximum likelihood decoder for
  medical images,''
\newblock {\em Multimedia Tools and Applications}, 2018.

\bibitem{chen2018quaternion}
Beijing Chen, Chunfei Zhou, Byeungwoo Jeon, Yuhui Zheng, and Jinwei Wang,
\newblock ``Quaternion discrete fractional random transform for color image
  adaptive watermarking,''
\newblock {\em Multimedia Tools and Applications}, 2018.

\bibitem{isac2011study}
Bibi Isac and V~Santhi,
\newblock ``A study on digital image and video watermarking schemes using
  neural networks,''
\newblock {\em International Journal of Computer Applications}, 2011.

\bibitem{jin2007applications}
Cong Jin and Shihui Wang,
\newblock ``Applications of a neural network to estimate watermark embedding
  strength,''
\newblock in {\em International Workshop on Image Analysis for Multimedia
  Interactive Services (WIAMIS)}, 2007.

\bibitem{kandi2017exploring}
Haribabu Kandi, Deepak Mishra, and Subrahmanyam RK~Sai Gorthi,
\newblock ``Exploring the learning capabilities of convolutional neural
  networks for robust image watermarking,''
\newblock {\em Computers \& Security}, 2017.

\bibitem{mun2017robust}
Seung-Min Mun, Seung-Hun Nam, Han-Ul Jang, Dongkyu Kim, and Heung-Kyu Lee,
\newblock ``A robust blind watermarking using convolutional neural network,''
\newblock {\em arXiv preprint arXiv:1704.03248}, 2017.

\bibitem{baluja2017hiding}
Shumeet Baluja,
\newblock ``Hiding images in plain sight: Deep steganography,''
\newblock in {\em Advances in Neural Information Processing Systems (NeurIPS)},
  2017.

\bibitem{shin2017jpeg}
Richard Shin and Dawn Song,
\newblock ``Jpeg-resistant adversarial images,''
\newblock in {\em NIPS 2017 Workshop on Machine Learning and Computer
  Security}, 2017.

\end{thebibliography}

\end{document}